\documentclass{jpp}
\usepackage{graphicx}

\usepackage[utf8]{inputenc}
\usepackage[T1]{fontenc}
\usepackage{amsmath}

\usepackage{color}

\newcommand{\bn}{\begin{equation}}
\newcommand{\bea}{\begin{eqnarray*}}
\newcommand{\eea}{\end{eqnarray*}}
\newcommand{\en}{\end{equation}}
\newcommand{\lang}{\left\langle}
\newcommand{\rang}{\right\rangle}

\newcommand{\vpar}{v_\parallel}

\newcommand{\simlt}{\:{\raisebox{-1.5mm}{$\stackrel
{\textstyle{<}}{\sim}$}}\:}

\shorttitle{Residual flow}
\shortauthor{G. G. Plunk and P. Helander}

\title{The residual flow in well-optimized stellarators}

\author{G. G. Plunk\aff{1}\corresp{\email{gplunk@ipp.mpg.de}},\and P. Helander\aff{1}}

\affiliation{\aff{1}Max-Planck-Institut für Plasmaphysik, 17491 Greifswald, Germany}

\begin{document}

\maketitle

\begin{abstract}
The gyrokinetic theory of the residual flow, in the electrostatic limit, is revisited, with optimized stellarators in mind.  We consider general initial conditions for the problem, and identify cases that lead to a non-zonal residual electrostatic potential, {\em i.e.} one having a significant component that varies within a flux surface.  We investigate the behavior of the ``intermediate residual'' in stellarators, a measure of the flow that remains after geodesic acoustic modes have damped away, but before the action of the slower damping that is caused by unconfined particle orbits.  The case of a quasi-isodynamic stellarator is identified as having a particularly large such residual, owing to the small orbit width achieved by optimization. 
\end{abstract}

\section{Introduction}

The work of Rosenbluth and Hinton \citep{Rosenbluth-Hinton-PhysRevLett.80.724, Hinton-Rosenbluth-1999} established the idea of an undamped ``residual'' potential in tokamaks.  The idea is to initialize an electrostatic potential, varying only in the radial direction, and track its value over a time much longer than the transit period over which particles move along the magnetic field lines.  Initially the potential oscillates and diminishes in amplitude, due to geodesic acoustic mode activity, but eventually a steady residual signal emerges.  Because all of this happens due to collisionless dynamics, it was argued that full gyrokinetics is needed to properly model the turbulence, which remains the prevailing attitude to this day.  

The residual proved very popular in part due to the fact that it admits exact predictions, commonly used to benchmark gyrokinetic codes.  However, the question of what such calculations imply for the behaviour of fully developed turbulence, where strongly nonlinear dynamics prevails, is a difficult one to grapple with.  Part of the difficulty is that the dynamics of the driven potential cannot be easily separated from that of the turbulence that drives it.  Indeed, modelling the turbulence as a steady source leads to unbounded growth of the residual \citep{Rosenbluth-Hinton-PhysRevLett.80.724}.  Other basic questions arise as to whether the residual is relevant in cases where the damped solutions called geodesic acoustic modes (GAMs) may be effectively driven by the turbulence \citep{waltz-holland}.

There is a vast body of work concerning zonal flows in magnetic fusion (see for instance \citet{Diamond_2005} for a start), with many simple limits having been considered theoretically, and a multitude of sometimes disparate models and explanations having been proposed.  There is however broad agreement that zonal flows have a beneficial influence on turbulence, lowering fluctuation levels by shearing turbulent eddies \citep{hahm_shear}, promoting transport of energy to small scales, and inducing coupling between unstable and stable eigenmodes \citep{Makwana_zfs}. Zonal flows are also responsible for the Dimits shift, whereby turbulence is all but eliminated for a finite range above the threshold of the linear instability \citep{Dimits_shift, Rogers-Dorland, st-onge_2017, hallenbert_plunk_2022, Pueschel_2021}.  In the context of stellarators, it is clear that the strength of the geodesic curvature, related to both GAM damping and residual levels, has a strong effect on the overall turbulence levels \citep{pavlos_zf}, and variation in the linear response of zonal flows is believed to underlie confinement differences between different configurations of the Large Helical Device (LHD) \citep{Watanabe-LHD-zf}.  The neoclassical radial electric field in a stellarator can also stabilize turbulence via shearing, {\em etc}, and is experimentally associated with enhanced confinement (see for instance \citep{Lore-HSX-eITB}), but its origins are distinct from zonal flows, and there is not yet evidence to support such a role in the W-7X stellarator \citep{pavlos2020}.  Overall there is a strong motivation to deepen the understanding of the theoretical foundations of zonal flows in stellarators, of which the residual is a key part, especially to aid in the design of future devices.

On the most fundamental theoretical level, one observes that, according to gyrokinetic theory, the entire `$k_\alpha  = 0$ component' of the fluctuations is stable (${\bf k}_\perp = k_\psi\bnabla \psi  + k_\alpha\bnabla \alpha $ is the wavenumber perpendicular to the magnetic field and ${\bf B} = \bnabla\psi\times\bnabla\alpha$ is the magnetic field), having no source of free energy.   That is, in addition to the zonal potential, which by definition is constant within a flux surface, there are also components that vary within a flux surface by virtue of smooth dependence along the field line, including the electrostatic potential and other moments like parallel ion flow or temperature perturbations.  What is the fate of these components, and do they also have a role in regulating the turbulence?

In the present work we focus on the residual in the context of stellarators instead of tokamaks (though we note that our findings apply equally to the latter).  The above issues also arise in a stellarator, and it is even more unclear for which cases the residual is a useful quantity as a predictor of the turbulence.  Indeed, as was found by \cite{Mishchenko-Helander-Koenies} and \cite{Helander_2011}, the residual in a stellarator is strongly affected by the presence of unconfined particle orbits, {\em i.e.} trapped particles bouncing back and forth in magnetic wells that drift radially between magnetic surfaces; see also \cite{Monreal-2016}.  If this drift is non-zero, however small, it was found that the residual is sharply reduced, an apparent strike against the stellarator.

However, in cases where the radial drift is sufficiently small, such as optimized stellarators, the additional stellarator-specific damping must necessarily act on a timescale much longer that the drive and saturation of the turbulence.  For such cases, we identify an `intermediate' residual, namely the solution that is found at times much longer than the transit time and much smaller than the timescale of the stellarator-specific damping.  To calculate this residual, we revisit the initial value problem of gyrokinetics, this time without assuming the initial condition or the final state to be well-approximated by a zonal potential.  Indeed, we find that the residual is generally non-zonal, but we do identify conditions under which such assumptions are valid.

This calculation casts the optimized stellarator in a more positive light, showing that it can exhibit a large residual, especially in the case that the width of particle orbits are small, which, as we demonstrate, is especially true in a particular class, so-called quasi-isodynamic stellarators. Indeed, in such stellarators the residual is found to be much larger than that in tokamaks, which could have significant consequences for the regulation of plasma turbulence. 

\section{Gyrokinetic solution of the initial value problem}

We will follow some of the notation conventions of \citet{Helander_2011}, but with some adaptations due to the fact that the derivation will use gyrokinetics instead of drift kinetics.  We are interested in solving the gyrokinetic system of equations in the electrostatic limit for the time evolution of a linear mode with wavenumber ${\bf k}_\perp = k_r \nabla r +  k_\alpha\nabla \alpha$, where $r(\psi)$ is an arbitrary radial coordinate that is constant on magnetic surfaces, and the magnetic field is expressed as ${\bf B} = \nabla\psi\times\nabla\alpha$ in terms of the toroidal flux function $\psi$ and Clebsch angle $\alpha$.  Here we are concerned only in the case $k_\alpha = 0$, where there is no source of free energy for the perturbation, so the linear mode is stable.  The collisionless gyrokinetic equation in this limit is

\begin{equation}
    \frac{\partial g_{a}}{\partial t} +
	v_{\|} \nabla_\| g_a + i \omega_{da} g_{a} 
=  \frac{e_a F_{a0}}{T_a} 
	\frac{\partial\phi}{\partial t} J_{0a} 
	\label{gk-eqn},
\end{equation}
    with $\nabla_\| = \hat{\bf b}\cdot\bnabla = \partial/\partial l$, where $\hat{\bf b} = {\bf B}/B$ and $l$ is the arc length along field line.  Here $g_a$ is the gyrocenter distribution function for species `$a$' and $F_{a0} = n_a (m_a/2\pi T_a)^{3/2} \exp(-m_a v^2/2 T_a)$, with $n_a$ and $T_a$ the bulk density and temperature.  The electrostatic potential $\phi$ is found from the quasi-neutrality constraint
\begin{equation}
    \sum_a n_a\frac{e_a^2}{T_a} \phi = \sum_a e_a \int g_{a} J_{0a} d^3v, 
	\label{field1}
\end{equation}
where $J_{na} = J_n(k_\perp v_\perp / \Omega_a)$, with $\Omega_a = e_a B/m_a$ and the velocity element is expressed in gyrokinetic phase space variables as
\begin{equation}
    d^3v = 2 \pi v_\perp dv_\perp v_\| = \sum_\sigma \frac{2 \pi B dE_a d\mu_a}{m_a^2 | v_\| |} = \sum_\sigma \frac{\pi B v^2 dv d\lambda}{\sqrt{1-\lambda B}},
\end{equation}
where we define $v_\perp = |\hat{\bf b}\times{\bf v}|$, $v_\| = \hat{\bf b}\cdot {\bf v}$, $E_a = m_a v^2/2$, $\mu_a = m_a v_\perp^2/(2 B)$, $v = |{\bf v}|$, and $\lambda = \mu_a/E_a$.  In what follows we mostly use phase space variables $v$ and $\lambda$, and express the parallel velocity as $v_\| = \sigma v \sqrt{1 - \lambda B}$ where $\sigma = v_\| / | v_\| | = \pm 1$.  Following \citet{Helander_2011}, the drift frequency is defined as $\omega_{da} = k_r {\bf v}_{da}\cdot\bnabla r$ with

\begin{equation}
    {\bf v}_{da}\cdot\bnabla r = \overline{v}_{ra} + \vpar \nabla_\|\delta_{ra},\label{eq:radial-drift-split}
\end{equation}
where $\overline{v}_{ra}$ denotes the transit averaged radial drift, which is zero for passing particles and, in the case of tokamaks and omnigenous stellarators, also for trapped particles.  The term $\vpar \nabla_\|\delta_{ra}$ is zero under orbit average, so $\delta_{ra}$ is the radial excursion, or ``orbit width'' of particles, a periodic function on the torus.  Eqn.~\ref{eq:radial-drift-split} simply represents the splitting of the radial drift into mean and oscillatory behavior with respect to the transit/bounce average.  The quantities $\overline{v}_{ra}$ and $\delta_{ra}$ are defined by this equation, and may be obtained as solutions of it, with the appropriate boundary condition for $\delta_{ra}$ in $l$; for this we take $\delta_{ra} = 0$ at bounce points of trapped-particle orbits, or at the field maximum for passing particles, implying that $\delta_{ra}$ is odd in $v_\|$.  The transit average, which is designed to annihilate the operator $v_\|\nabla_\|$, is defined as

\begin{equation}
    \overline{f} = \frac{1}{2}\sum_\sigma\int^{l_2}_{l_1} \frac{f}{\sqrt{1 - \lambda B(l)}}dl\bigg\slash\int^{l_2}_{l_1} \frac{1}{\sqrt{1 - \lambda B(l)}}dl,
\end{equation}
where, for trapped particles, $l_1$ and $l_2$ are the bounce points (where $v_\| = 0$) such that $B(l_1)=B(l_2) = 1/\lambda$, while for passing particles ($\lambda < 1/B_\mathrm{max}$) the average is understood in the limiting sense, as $l_1 \rightarrow -\infty$ and $l_2 \rightarrow \infty$.\footnote{The factor of $1/2$ is chosen so that the summation is evaluated $\frac{1}{2}\sum_\sigma = 1$ for the typical case when $f$ is independent of $\sigma$.  Note that transit averaged quantities depend on $\lambda$ and also the "well" index denoted $j$.  We reserve the zeroth index $j = 0$ to denote the unbounded domain of passing particles; see appendix \ref{appx:identities}.}  We also define the flux surface average
\citep{helander2014theory},
\begin{equation}
    \lang \cdots \rang = \lim_{L\rightarrow \infty} \int_{-L}^L (\cdots ) \frac{dl}{B} \bigg\slash
	\int_{-L}^L \frac{dl}{B}.
\end{equation}

Following previous works we take the Laplace transform of Eqn.~\ref{gk-eqn}, defining $\hat{g}_a = \int_0^{\infty} dt \exp(-pt) g_a(t)$, and introduce an integrating factor to absorb the orbit width term, defining $h_a = \exp(i k_r \delta_{ra}) g_a$

\begin{equation}
    (p + i k_r \overline{v}_{ra})\hat{h}_a + v_\| \nabla_\| \hat{h}_a = \left[p \frac{e_a\hat{\phi}}{T_a} J_{0a} F_{a0}  + \delta F_a(0)\right] e^{i k_r \delta_{ra}},
\end{equation}
where $\delta F_a(0)$ is the initial value of the gyrocentre distribution function,

\begin{equation}
    \delta F_{a} = g_{a} - \frac{e_a \phi}{T_a}J_{0a} F_{a0}, 
	\label{eq:delta-F}
\end{equation}
We are interested in times much longer that the transit/bounce timescale $\omega_b \hat{h} \sim v_\| \nabla_\| \hat{h}$, which we order similar to the non-secular part of the drift frequency, {\em i.e.} $\omega_b \sim k_r v_\|\nabla_\|\delta_r$, but $p \sim k_r \overline{v}_r \ll \omega_b$.  This implies that at dominant order we have $\partial \hat{h}/\partial l = 0$.  At next order, we transit average and use continuity of $g_a$ at bounce points, which, because $\delta_{ra} = 0$ at such points, implies $h_a|_{\sigma=1}=h_a|_{\sigma=-1}$ at all $l$, yielding

 \begin{equation}
     (p + i k_r \overline{v}_{ra}) \hat{h}_a = \left(p\overline{\frac{e_a\hat{\phi}}{T_a} J_{0a}e^{i k_r \delta_{ra}}}F_{a0} + \overline{\delta F_a(0) e^{i k_r \delta_{ra}}}\right).\label{eq:gk-asymp-soln-h}
 \end{equation}
The solution for $\hat{g}_a$ is therefore

\begin{equation}
    \hat{g}_a = \frac{1}{p + i k_r \overline{v}_{ra}} \left(p\overline{\frac{e_a\hat{\phi}}{T_a} J_{0a}e^{i k_r \delta_{ra}}}F_{a0} + \overline{\delta F_a(0) e^{i k_r \delta_{ra}}}\right)e^{-i k_r \delta_{ra}}.\label{eq:gk-asymp-soln-g}
\end{equation}
This is the same as the result of \citet{Helander_2011} except that it retains the gyro-average ($J_{0a}$) and keeps $\phi$ and $\delta F_a$ under the orbit average, allowing them to vary within the flux surface.  A similar comparison can be made to the result of \citet{Monreal-2016}, which also retains the full gyro-average but not the mentioned flux surface dependence.  To obtain an equation simply for $\phi$ we substitute this expression into the quasi-neutrality condition \ref{field1}, which gives

\begin{multline}
    \sum_a \frac{e_a^2}{T_a}\left(n_a \hat{\phi} - \int d^3v J_{0a}F_{a0} \frac{p}{p + i k_r \overline{v}_{ra}}\overline{\hat{\phi} J_{0a}e^{i k_r \delta_{ra}}} e^{-i k_r \delta_{ra}}\right)\\
    = \sum_a e_a \int d^3v J_{0a}\frac{1}{p + i k_r \overline{v}_{ra}}\overline{\delta F_a(0) e^{i k_r \delta_{ra}}}e^{-i k_r \delta_{ra}}.\label{eq:full-integral-eqn}
\end{multline}

Solving this equation, which can be compared with Eqn.~(6) of \citet{Helander_2011} (see also appendix \ref{appx:zf-oscillations}), would give the fully general solution for the post-GAM zonal flow dynamics, allowing for arbitrary orbit widths ($k \delta_r \sim k\rho \sim 1$).  One would like to be able to solve Equation \ref{eq:full-integral-eqn} for $\hat{\phi}$ then invert the Laplace transform and obtain $\phi(t)$.  Unfortunately the situation is rather complicated, as $\hat{\phi}$ appears under an orbit average involving a resonant velocity integral, so we will have to take some limits to make further progress.

\subsection{Limit of small orbit width and ion Larmor radius}

Following previous works, we now consider the limit 

\begin{equation}
    k_r\delta_{ra} \sim k_\perp\rho_a \ll 1.
\end{equation}
We will later take $k_\perp \rho_e = k_r \delta_{re} = 0$, since $\rho_e \ll \rho_i$ and $\delta_{re} \ll \delta_{ri}$.  Some care is needed here as the polarization effects arising in the gyrokinetic equation enter at the same order as those which appear in the quasi-neutrality constraint, which itself is singular.  This is made clear by recasting the quasi-neutrality condition \ref{field1} in terms of the gyrocentre distribution function,

\begin{equation}
    \sum_a n_a \frac{e_a^2}{T_a} \left[1 - \Gamma_0(b_a) \right] \phi 
	= \sum_a e_a \int \delta F_{a} J_{0a} d^3v = \sum_a e_a \delta n_a, 
	\label{field4}
\end{equation}
where $\Gamma_0(x) = I_0(x) e^{-x}$ and $b_a = k_\perp^2 \rho_a^2 = k_\perp^2 T_a / (m_a \Omega_a^2)$.  Note that in the final expression we introduce the ``gyrocenter density'' $\delta n_a$.  The point is that the gyrocenter density is small in the limit $b_i \ll 1$, {\em i.e.}, taking $\Gamma_0(b) \approx 1 - b$, one sees that $\delta n/n \sim {\cal O}(b_i e_i\phi/T_i)$,
\begin{equation}
    \sum_a e_a \delta n_a =  b_i n_i \frac{e_i^2 \phi}{T_i}
\end{equation}
where $b_i = k_\perp^2 \rho_i^2 = k_\perp^2 m_i T_i/(e_i^2B^2)$, which means that it is necessary to include terms of order $b_i$ to solve for the electrostatic potential in this limit.  

We now perform the expansion on Eqn.~\ref{eq:full-integral-eqn}, ordering $b_i e_i \phi/T_i \sim \delta n_a/n_a$.  To put the final expression in a more convenient form, we rewrite the resonant term on the left hand side using $p/(p + i k_r \overline{v}_{ra}) = 1 - i k_r \overline{v}_{ra}/(p + i k_r \overline{v}_{ra})$, and thus obtain

\begin{multline}
\sum_a n_a\frac{e_a^2}{T_a}\left(\hat{\phi}  - \frac{1}{n_a} \int d^3v F_{a0} \overline{\hat{\phi}}\right) + \sum_a \frac{e_a^2}{T_a}\int d^3 v \frac{i k_r \overline{v}_{ra}}{p + i k_r \overline{v}_{ra}}F_{a0}\overline{\hat{\phi}} \\
+ \sum_a \frac{e_a^2}{T_a}\int d^3 v F_{a0}\left[\overline{\hat{\phi}} b_a x_\perp^2/2 + \overline{\hat{\phi} b_a x_\perp^2/2} - k_r^2 \overline{\hat{\phi}\delta_{ra}}\delta_{ra} + k_r^2 \overline{\hat{\phi}\delta_{ra}^2}/2+ k_r^2 \overline{\hat{\phi}}\delta_{ra}^2/2\right] \\
= \frac{1}{p}\sum_a e_a \int d^3v J_{0a}\left(1 - \frac{i k_r \overline{v}_{ra}}{p + i k_r \overline{v}_{ra}}\right)\overline{\delta F_a(0) e^{i k_r \delta_{ra}}}e^{-i k_r \delta_{ra}},\label{eq:full-integral-eqn-2}
\end{multline}
where we have written $x_\perp^2 = m_a v_\perp^2 / (2 T_a)$, expanded the Bessel function as $J_{0a} (k_\perp v_\perp / \Omega_a) \simeq 1 - b_a x_\perp^2 / 2$, and recognised that terms that are linear in $\delta_{ra}$ are odd in $v_\|$ and vanish upon integration. This equation is still fairly complicated, but the terms can be identified.  On the left hand side, the first term shows contributions from the non-zonal part of the potential: it is zero under zonal average and also for the case $\phi = \left<\phi\right>$; we return to this later.  The second term contains the resonance that yields zonal flow oscillations and damping in stellarators with $\overline{v}_{ra} \neq 0$ \citep{Mishchenko-Helander-Koenies,Helander_2011,Monreal_2017}.  This is not the focus of the present work, although we discuss it briefly in Appendix \ref{appx:zf-oscillations}.  Note that the finite orbit effects on this term are neglected, which is justified in the limit of small $k_r\overline{v}_{ra}/p$.  On the second line we encounter all the finite orbit width (FOW) and finite Larmor radius (FLR) terms associated with the residual.  These expressions will simplify significantly (and become more familiar) in limits when the potential is mostly zonal.  Finally, on the right hand side we have the contribution from the initial condition; note the separation into a resonant term and another which can be evaluated using quasi-neutrality in terms of the initial potential.  We keep this term exact, for now, since we would like to discuss the consequences of several possible orderings for the initial condition itself, $\delta F_a(0)$ in the following section; see also Appendix \ref{appx:deltaF0}.

\section{The residual potential}

Let us consider the limit where the resonance can be neglected, {\em i.e.} let us take $p \gg k_r \overline{v}_{ra}$.  Here, we neglect damping special to stellarators, which includes both exponential and algebraic decay \citep{Helander_2011}.  Although that damping process is slow, it can still manage to deplete most of the zonal amplitude.  Indeed, as shown by Eqn.~(16)-(17) of \citet{Helander_2011}, the final residual is independent of the size of the damping rate when that rate is small, {\em i.e.} any non-zero value of $\overline{v}_{ra}$ results in a strong correction to the residual that depends on the fraction of trapped particles.  Therefore the `residual' that arises before this decay takes effect may be more relevant for understanding the interaction between zonal flows and turbulence, especially in optimized stellarators where $\overline{v}_{ra}$ is made to be as close to zero as possible.  We therefore focus on the ``intermediate residual'' defined to be the value of the potential long after the GAMs have decayed, $\tau_\mathrm{G} \sim 1/\gamma_{\mathrm{GAM}}$, but long before the final residual is obtained, on timescale of the stellarator-specific damping of Mishchenko and Helander, $\tau_\mathrm{M} \sim 1/(k_r \overline{v}_{ra})$:

\begin{equation}
    \phi_\mathrm{res} \equiv \lim_{\frac{\tau_\mathrm{M}}{t} \rightarrow \infty} \left (\lim_{\frac{t}{\tau_\mathrm{G}} \rightarrow \infty}  \phi(t) \right)\label{eq:residual-def}
\end{equation}
For tokamaks (and perfectly omnigenous stellarators) this quantity coincides with the conventional definition of the residual, as defined by \citet{Rosenbluth-Hinton-PhysRevLett.80.724}, as shown in what follows.  

To be slightly more formal, we note that the inner limit of Eqn.~\ref{eq:residual-def} has already been taken much earlier in our calculation to obtain Eqn.~\ref{eq:gk-asymp-soln-h}, and additional limits are to be considered subsidiary to that one.  In particular $(t \omega_b)^{-1} \ll k_r \overline{v}_{r} t \ll 1$, with the latter condition expressing the outer limit of Eqn.~\ref{eq:residual-def}.  According to these orderings, $t$ is assumed to be both large and small, {\em i.e.} $t \gg \omega_b^{-1}$ but $t \ll (k_r \overline{v}_{r})^{-1}$ (implying $k_r \overline{v}_{r}/\omega_b \ll 1$), which, because $\overline{v}_{r} = 0$ is never exactly true in an actual stellarator, can only ever approximately be satisfied within a finite time interval, between $\omega_b^{-1}$ and $(k_r \overline{v}_{r})^{-1}$.  The observation of $\phi_\mathrm{res}$ may therefore be a challenge in some cases, for instance in gyrokinetic simulations of stellarators for which these timescales are not well separated.

Obtaining the desired limit in our calculation is however a simpler matter, as we need only apply $k_r \overline{v}_{ra}/p \rightarrow 0$ to Eqn.~\ref{eq:full-integral-eqn-2}.  The only remaining dependence on $p$ is the factor of $1/p$ on the source term, and the Laplace transform may be inverted to obtain

\begin{multline}
    \sum_a \frac{e_a^2}{T_a}\int d^3 v F_{a0}\left[\overline{\phi} b_a x_\perp^2/2 + \overline{\phi b_a x_\perp^2/2} - k_r^2 \overline{\phi\delta_{ra}}\delta_{ra} + k_r^2 \overline{\phi\delta_{ra}^2}/2 + k_r^2 \overline{\phi}\delta_{ra}^2/2\right]\\
    + \sum_a n_a\frac{e_a^2}{T_a}\left(\phi  - \frac{1}{n_a} \int d^3v F_{a0} \overline{\phi}\right) = S,\label{eq:full-integral-eqn-3}
\end{multline}
where the right-hand side denotes the source term (not yet expanded for small orbit width),
\begin{equation}
    S = \sum_a e_a\int d^3v J_{0a}\overline{\delta F_a(0) e^{i k_r \delta_{ra}}}e^{-i k_r \delta_{ra}}.
\end{equation}
Eqn.~\ref{eq:full-integral-eqn-2} can be compared with previous results in the small orbit width and small Larmor radius limits (for completeness, the result valid for arbitrary $k^2\delta_r^2$ and $b_i$ is given in Appendix \ref{appx:general-residual}).  We note differences coming from the fact that potential is kept under the bounce average, because we allow for $\phi \neq \left<\phi\right>$, and the finite Larmor radius terms take a similar form.

Let us consider what can be said about the general solution of equation \ref{eq:full-integral-eqn-2}.  The final term on the left hand side of Eqn.~\ref{eq:full-integral-eqn-2} has the form of a linear operator on $\phi$, defined by

\begin{equation}
    \sum_a n_a\frac{e_a^2}{T_a}\left(\phi  - \frac{1}{n_a} \int d^3v F_{a0} \overline{\phi}\right) = \sum_a n_a\frac{e_a^2}{T_a} {\cal L} \phi ,
\end{equation}
which is zero if and only if $\phi = \left<\phi\right>$; see Appendix \ref{appx:positivity}.  As a consequence, this operator is invertible on the non-zonal part of the potential, $\delta\phi$, defined by the following

\begin{equation}
    \phi = \delta\phi + \Phi,
\end{equation}
where $\Phi = \left<\phi\right>$.  We may formally expand both $\Phi = \Phi^{(0)} + \epsilon \Phi^{(1)} + \dots$, $\delta \phi = \delta \phi^{(0)} + \epsilon \delta\phi^{(1)} + \dots$ and $\delta F_a = \delta F_{a}^{(0)} + \dots $ in our small parameter ($\epsilon \sim b_i \sim k_r^2\delta_{ri}^2$); it will not be necessary to keep these extra superscripts in what follows because we will only use zeroth quantities in our final expressions.  With this expansion, the dominant contributions to Eqn.~\ref{eq:full-integral-eqn-2} come from the right-hand side and the term $\sum_a n_a\frac{e_a^2}{T_a} {\cal L} \phi$ on the left-hand-side.  The resulting equation can be formally solved for $\delta\phi^{(0)}$ by use of the inverse ${\cal L}^{-1}$, yielding the non-zonal part of the residual potential,

\begin{equation}
    \delta\phi_\mathrm{res} = \left(\sum_a n_a\frac{e_a^2}{T_a}\right)^{-1} {\cal L}^{-1} \left[S_0 - \left<S_0\right>\right],\label{eq:non-zonal-residual}
\end{equation}
where $S_0 = \sum_a e_a\int d^3v \overline{\delta F_{a}^{(0)}(0)}$.  This equation implies that the residual potential essentially derives its non-zonal component from non-uniformity of initial charge distribution on the surface (or to be more precise, the charge density of the transit average of the gyro-average of the initial distribution functions).  The main conclusion here, which may or may not be surprising, is that this component does not in fact decay away to zero.

At next order in our expansion, equation \ref{eq:full-integral-eqn-2}, the contributions from $\delta\phi^{(1)}$ appearing under the operator ${\cal L}$ are eliminated by flux-surface average, leaving an equation for the zonal part of the potential, $\Phi^{(0)}$:

\begin{multline}
\Phi_\mathrm{res} = \\
\frac{\left<b_i \phi(0)\right> +\widetilde{\Phi}_S - (2n_i)^{-1}\left<\int d^3 v F_{i0}\left[\overline{\delta\phi_{res}} b_i x_\perp^2 + \overline{\delta\phi_{res} b_i x_\perp^2} + k_r^2 \overline{\delta\phi_{res}\delta_{r}^2} + k_r^2 \overline{\delta\phi_{res}}\delta_{r}^2\right]\right>}{\left<b_i \right> +  n_i^{-1}\left<\int d^3v F_{i0}k_r^2\delta_{r}^2 \right>},\label{eq:full-residual}
\end{multline}
where we use $\delta_{re} \ll \delta_{ri} \equiv \delta_{r}$ (dropping indices), $\rho_e \ll \rho_i$, the identity $\left< \overline{f}\right> = \left< f \right>$ (equation \ref{eq:flux-bounce-avg}), $\overline{\delta_{ra}} = 0$ (due to oddness in $\sigma$ for trapped particles, and by choice of convention for passing particles) and quasi-neutrality for the initial condition.  The latter can be written at each order in terms of the the $n$-th order initial condition $\delta F_{a}^{(n)}(0)$, but these details are left for Appendix \ref{appx:deltaF0}.  Additional contributions from the source at this order are included in the term
\begin{equation}
    \widetilde{\Phi}_S = \frac{T_i}{2n_ie_i} \left<\int d^3v \left(2i \overline{k_r \delta_{r}\delta F_i(0)} + \left(\delta F_i(0)-\overline{\delta F_i(0)}\right) b_i x_\perp^2 - k_r^2 \overline{\delta F_i(0)\delta_{r}^2} - k_r^2 \overline{\delta F_i(0)}\delta_{r}^2 \right)\right>.
\end{equation}
Note the mixed orders of the terms: Here it is possible to consider, just for example, initial conditions $\delta F_i(0) \sim {\cal O}(\delta_r k_r)$ that are odd in $v_\|$, as done by \citet{Rosenbluth-Hinton-PhysRevLett.80.724}, or even contributions at order $\delta F_i(0) \sim {\cal O}(1)$, for instance due to pressure perturbations.  However, we emphasize that such cases are generally inconsistent with the assumption that the residual is zonal, and the rather unwieldy expression of equation \ref{eq:full-residual} must then be considered to determine the residual.  The conditions under which this general solution prevails depend on details of the turbulence, and this will be discussed more later.

We note that a closed form expression for the zonal part of the residual ($\Phi_\mathrm{res}$) in terms of the source may be obtained by substituting the solution for $\delta\phi_\mathrm{res}$, equation \ref{eq:non-zonal-residual}, into equation \ref{eq:full-residual}; we do not do this here as the resulting expression is not enlightening, especially without an explicit form of ${\cal L}^{-1}$.

\subsection{Recovering the residual zonal flow}
If the contribution to the charge from the initial gyro-center distribution, {\em i.e.} the right hand size of equation \ref{eq:full-integral-eqn-2}, is constant on a flux surface (to zeroth order), then the $\delta \phi$ term must balance with the small (FLR, FOW) terms, and we can conclude

\begin{equation}
    \delta\phi \sim {\cal O}(b_i \Phi),\label{eq:nz-ordering}
\end{equation}
and $\delta\phi$ can be safely neglected in equation \ref{eq:full-residual}; those from $\widetilde{\Phi}_S$ can also be neglected if $\delta F_i(0) \sim {\cal O}(b_i)$ is assumed.  Solving the equation for $\Phi$ we then obtain the residual

\begin{equation}
    \phi_\mathrm{res} = \frac{\left<b_i \phi(0)\right>}{\left<b_i \right> +  n_i^{-1}\left<\int d^3v F_{i0}k_r^2\delta_{r}^2 \right>}.\label{eq:res-zf}
\end{equation}
We see that, even in this limit, we do not exactly recover the result of \cite{Rosenbluth-Hinton-PhysRevLett.80.724}, as we do not assume the initial potential to be zonal.  The RH result can be written in our notation as follows:
\begin{equation}
    \phi_\mathrm{res}^\mathrm{RH} = \Phi(0)\frac{\left<b_i\right>}{\left< b_i\right> + n_{i}^{-1}\left<\int d^3v F_{i0}k_r^2\delta_r^2 \right>},\label{eq:res-zf-RH}
\end{equation}
where we have used that \citet{Rosenbluth-Hinton-PhysRevLett.80.724} assumed the initial potential to be zonal to write this result in terms of $\phi(0) = \left<\phi(0)\right> = \Phi(0)$.  Evidently, there is one limit (for arbitrary $\phi(0)$) in which the Rosenbluth-Hinton (RH) result is obtained from equation \ref{eq:res-zf}, which is when $\left< b_i \right> \approx b_i$, {\it e.g.} for the circular tokamak model.

\subsection{Dependence of residual zonal flow on initial potential}

More generally, we note that there is a class of initial conditions consistent with equation \ref{eq:nz-ordering}, including that traditionally assumed in calculations of the residual, $\phi(0) = \left<\phi(0)\right> = \Phi(0)$.  Indeed, allowing non-zonal $\phi(0)$, equation \ref{eq:res-zf} exhibits a certain variation in what can be obtained for the ratio $\phi_\mathrm{res}/\Phi(0)$, as compared to the RH expression, {\em i.e.}

\begin{equation}
    \frac{\phi_\mathrm{res}}{\phi_\mathrm{res}^\mathrm{RH}} = \frac{\left<b_i \phi(0)\right>}{\left<b_i\right>\left<\phi(0)\right>},
\end{equation}
which arises only from the variation in $b_i$.  An interesting and possibly useful case is that of initially zonal distribution functions (a convenient way to initialize a gyrokinetic simulation and therefore a good test case).  In this case the initial charge is also zonal and from equation \ref{eq:initial-charge} we have

\begin{equation}
    \left<b_i \phi(0)\right> = b_i \phi(0).\label{eq:res-zf-criterion}
\end{equation}
Dividing by $b_i$ and averaging we can solve for this for the zonal potential, $\Phi(0) = \left<\phi(0)\right>$, and obtain 
\begin{equation}
    \phi_\mathrm{res} = \Phi(0) \frac{\left<b_i^{-1} \right>^{-1}}{\left<b_i \right> +  n_i^{-1}\left<\int d^3v F_{i0}k_r^2\delta_{r}^2 \right>}.\label{eq:res-zf-2}
\end{equation}
Because of the inequality between the harmonic and arithmetic means, we find that the residual expressed by \ref{eq:res-zf-2} is less than or equal to the RH expression, in particular $\phi_\mathrm{res}/\Phi(0) \leq \phi_\mathrm{res}^\mathrm{RH}/\Phi(0)$, with equality only in the case of uniform $b_i$.

\section{The residual in stellarators and tokamaks}

Having demonstrated how to calculate the intermediate residual for stellarators, the natural question arises about how different stellarators fare with respect to this measure, how they compare with tokamak, and in particular whether anything can be said about the different classes of optimized stellarators.

\subsection{Tokamaks and quasi-symmetric stellarators}

The residual is inversely proportional to a weighted average of $\delta_r^2$ and will thus be particularly large in a field where the radial width of most particle orbits is small. In a standard large-aspect-ratio tokamak with circular cross section -- the case considered by \citet{Rosenbluth-Hinton-PhysRevLett.80.724}  -- circulating ion orbits have radial excursions of order $q \rho_i$ whereas trapped ones have larger banana orbits of width
    \begin{equation} 
    \delta_r \sim \frac{q \rho_i}{\epsilon^{1/2}},
    \end{equation}
where $\epsilon \ll 1$ denotes the inverse aspect ratio and $q=\iota^{-1}$ the inverse rotational transform \citep{Helander-Sigmar}. Although the latter only constitute a small fraction $f_t \sim \epsilon^{1/2} \ll 1$ of the total number of particles, they dominate the average of $\delta_r^2$, which becomes of order
    \begin{equation} \ \frac{1}{n_i} \int d^3v F_{i0} \delta_r^2 \sim (1-f_t) q^2 \rho_i^2 + f_t \left( \frac{q \rho_i}{\epsilon^{1/2}} \right)^2 
    \sim \frac{q^2 \rho_i^2 }{\epsilon^{1/2}},
        \end{equation}
and qualitatively explains the RH result
    \begin{equation} 
    \phi_\mathrm{res}^\mathrm{RH} = \frac{\Phi(0)}{1 + 1.64 q^2\epsilon^{-1/2}}.
    \label{tokamak response}
    \end{equation}
Since, in a typical tokamak, the term $1.64 q^2\epsilon^{-1/2}$ is considerably larger than unity, this residual is relatively weak. 
        
In quasisymmetric stellarators, particle trajectories are similar those in tokamaks \citep{Boozer-isomorphism,Nuhrenberg-Zille}, and the calculation is therefore mathematically identical if the symbols are suitably re-interpreted. In quasi-axisymmetric stellarators, the orbit width is equal to that in a tokamak, and the residual is therefore given by an expression like Eqn.~(\ref{tokamak response}), except that the numerical factor 1.64 needs to be adjusted if the magnetic field strength does not vary sinusoidally along the field. A similar adjustment is required in shaped tokamaks \citep{Xiao-Catto}. In quasihelically symmetric stellarators, the banana orbit width,
    \begin{equation}
        \delta_r \sim \frac{\rho_i}{|N - \iota|},
    \end{equation}
is smaller than that in a tokamak by a factor $|N/\iota - 1|>1$. Such stellarators thus have a larger residual than quasi-axisymmetric ones and tokamaks. 

\subsection{Quasi-isodynamic stellarators}

The smallest orbit widths, and thus the largest residuals, are realised in so-called quasi-isodynamic stellarators\footnote{Zero orbit width is theoretically achieved in an isodynamic magnetic field, which is however impossible to realise in practice \citep{helander2014theory}.}, which are omnigenous stellarators with poloidally closed contours of constant field strength. Such stellarators usually do not carry any significant amount of net toroidal current, and the magnetic field can be written as \citep{helander2014theory}
    \begin{equation}
        {\bf B} = G(\psi) \nabla \varphi + K(\psi, \alpha, \varphi) \nabla \psi
    \end{equation}
in Boozer coordinates, and the radial drift velocity becomes
    \begin{equation}
        v_r = \frac{v_\|^2 + v_\perp^2/2}{\Omega} ({\bf b} \times \nabla \ln B) \cdot \nabla r =
        - \frac{v^2 r'(\psi)}{\Omega}  \left(1 - \frac{\lambda B}{2} \right) 
        \left( \frac{\partial B}{\partial \alpha} \right)_{\psi,\varphi}, 
        \label{eq:vr}
    \end{equation}
where $\Omega = eB/m$ denotes the gyrofrequency.  Eqn.~\ref{eq:radial-drift-split} can be solved for the radial excursion
    \begin{equation}
        \delta_r = \frac{1}{v} \int v_r dt, 
    \end{equation}
where we use $\overline{v}_r = 0$ and define $dt = dl/\sqrt{1-\lambda B}$ with $t$ a time-like variable along the orbit.  The lower limit of integration can be chosen so that $\overline{\delta_r} = 0$. For magnetically trapped orbits, i.e. for values of $\lambda$ less than $1/B_{\rm max}$, this is achieved by choosing the lower integration limit to correspond to a bounce point and for passing orbits to the point of maximum field strength. In the latter case, 
    \begin{equation} 
    \overline{\delta_r} = \lim_{L\rightarrow \infty} \int_0^L \frac{\delta_r dl}{\sqrt{1-\lambda B}} \bigg\slash
	\int_0^L \frac{dl}{\sqrt{1-\lambda B}} = \left\langle \frac{B \delta_r}{\sqrt{1-\lambda B}} \right\rangle \bigg\slash
    \left\langle \frac{B}{\sqrt{1-\lambda B}} \right\rangle
 \end{equation}
 vanishes thanks to the $\alpha$-derivative in Eqn.~(\ref{eq:vr}), because the flux-surface average can be written as \citep{Helander_2009}
    \begin{equation}
        \langle \cdots \rangle = \frac{1}{V'} \int_0^{2\pi} d\alpha \int_0^L\frac{dl}{B} (\cdots ) ,
    \end{equation}
where 
    \begin{equation}
        V'(\psi) = \int_0^{2\pi} d\alpha \int_0^L \frac{dl}{B}.
    \end{equation}
Here, integral over the arc length $l$ is taken over one period of the device from one maximum of $B$ to the next, and the distance $L$ between them is independent of $\alpha$, which is true for perfectly quasi-isodynamic fields.

We shall not endeavour to calculate the zonal-flow response (\ref{eq:res-zf}) explicitly but rather show that it is relatively large in quasi-isodynamic stellarators by finding an upper bound on the quantity 
    \begin{equation}
        D = \frac{1}{n} \left\langle \int d^3v F_{0} \delta_r^2 \right\rangle 
        = \frac{1}{nV'} \int_0^{2\pi} d\alpha \int_0^L \frac{dl}{B} \int_0^\infty F_0 2 \pi v^2 dv \int_0^{1/B} \frac{\delta_r^2 B d\lambda}{\sqrt{1-\lambda B}}.
    \end{equation}
By interchanging the integrals over $\lambda$ and $l$, and replacing the latter by $t$, we find
    \begin{equation}
        D = \frac{1}{nV'} \int_0^\infty F_0 2 \pi v^2 dv \int_0^{2\pi} d\alpha \int_0^{1/B_{\rm min}} d \lambda \int \delta_r^2 dt,
    \end{equation}
where the $t$-integral is taken over the region where $1-\lambda B $ is positive, i.e. over the entire range for passing orbits and over the magnetic trapping well(s) for trapped ones. At the end points of each $t$-integral, the function $\delta_r$ vanishes. As a consequence of the Poincar{\'e} inequality,
    \begin{equation}
        \int_0^{\tau_b} g^2(t) dt \le \frac{\tau_b^2}{\pi^2} \int_0^{\tau_b} \left( \frac{dg}{dt}\right)^2 dt, 
    \end{equation}
for functions such that $g(0) = g(\tau_b) = 0$, we thus conclude that $D$ is bounded from above by
   \begin{equation}
        D \le \frac{2}{\pi nV'} \int_0^\infty F_0 dv \int_0^{2\pi} d\alpha \int_0^{1/B_{\rm min}} \tau_b^2
        d \lambda \int v_r^2 dt,
    \end{equation}
where 
     \begin{equation}
     \tau_b(\lambda) = \int_{\lambda B(l) < 1} \frac{dl}{\sqrt{1- \lambda B(l)}}
     \label{bounce time}
     \end{equation}
Substituting Eqn.~(\ref{eq:vr}) finally results in the rigorous inequality 
    \begin{equation}
        D \le \frac{3 m T}{2 \pi^2 e^2 V'} \left( \frac{dr}{d \psi} \right)^2 \int_0^{2\pi} d\alpha
         \int_0^L \left( \frac{\partial \ln B}{\partial \alpha} \right)_{\psi,\varphi}^2 dl
        \int_0^{1/B} \tau_b^2 
        \left( 1 - \frac{\lambda B}{2} \right)^2 \frac{d\lambda}{\sqrt{1-\lambda B}}.\label{eq:D-inequality-1}
    \end{equation}
The integrals in this expression depend on details in the spatial variation in the magnetic field strength, but we note that generally the $\lambda$-integral is of order $L^2/B$ and the $l$-integral of order $\epsilon^2 L$, where $\epsilon$ denotes the relative poloidal variation of $B$ at constant $\varphi$. Since $d \psi / dr \sim rB$ and $V' \sim 2 \pi L/B$, we thus obtain 
    \begin{equation}
        D \simlt \frac{3}{2 \pi^2} \left( \frac{\epsilon \rho_i L}{r} \right)^2.
        \label{eq:estimate for D}
        \end{equation}
In a quasi-isodynamic stellarator, the level-curves of constant magnetic field strength close poloidally, rather than toroidally, on each flux surface. The field strength varies along the magnetic axis, where it is a function only of $\varphi$, and in its vicintiy, the quantity $\epsilon = \partial \ln B / \partial \alpha = \partial \ln B / \partial \theta$ appearing in Eqn.~\ref{eq:estimate for D} is thus small. In the typical large-aspect-ratio scenario, we can estimate $\epsilon \sim r\kappa$ where $\kappa$ is the curvature of the magnetic axis; see for example \cite{plunk_landreman_helander_2019}.  A conservative bound for this curvature is $\kappa \simlt 1/L$ (optimization may achieve somewhat lower values), yielding $\epsilon \simlt r/L$.  We therefore expect from Eqn.~(\ref{eq:estimate for D}) that $D \simlt \rho_i^2$ for a quasi-isodynamic stellarator, and the residual (\ref{eq:res-zf}) is therefore comparable to the initial perturbation, i.e. much larger than in a tokamak. 
        

\section{Conclusions}

Although the long-time asymptotic residual potential is expected to be small in a stellarator \citep{Mishchenko-Helander-Koenies, Helander_2011}, we have argued that a well-optimized stellarator exhibits a larger effective residual on timescales important for turbulence.  To assess this `intermediate' residual, we have revisited the general initial value problem, allowing for arbitrary initial condition, and derive the resulting form of the residual, whether zonal or non-zonal.  We identify two cases (in the limit of small orbit width and Larmor radius) depending on the charge induced by the double-orbit-averaged (bounce- and gyro-averaged) initial distribution functions, which can be described as follows.  

If this charge is zonal (constant on a flux surface), we find that the residual potential is also zonal, and depends only on the initial potential, {\i.e.} it is insensitive to other details of the initial distribution functions.  In that case, we note that a large `intermediate' residual is indeed possible in stellarators, even in cases when the `true' time-asymptotic residual is negligibly small.  It is argued that the intermediate residual will be largest in quasi-isodynamic stellarators, smaller in quasi-helically symmetric stellarators, and smallest in tokamaks and quasi-axisymmetric stellarators.  This may be counter-intuitive to some readers, as it is known that undamped equilibrium flows can be sustained on time scales exceeding the ion collision time in quasi-symmetric stellarators but not in quasi-isodynamic ones \citep{Helander-Simakov-PRL,Helander-Geiger-Maassberg}, but there is a distinction between those equilibrium flows and the small-scale zonal flows that arise spontaneously with micro-turbulence.  For the latter flows, which regulate turbulence at low collisionality, it is the quasi-isodynamic stellarator that performs the best. These stellarators exhibit a much larger residual than tokamaks and quasi-axisymmetric stellarators where the Rosenbluth-Hinton factor $1.6 q^2 \epsilon^{-1/2}$ substantially exceeds unity. The collisional damping that occurs in quasi-isodynamic fields (but not in quasisymmetric ones) due to the lack of intrinsic ambipolarity only takes place on the longer time scale of ion collisions.

Formal complications arise in the calculation of the residual when the charge induced by the initial condition has a significant non-zonal component.  We show in this case that the residual potential is non-zonal, {\em i.e.} varies in the flux surface, and generally depends on details of the initial distribution functions.  We work out the general form of the complete solution, leaving its more detailed analysis for later, but note that its understanding may allow the consideration of a broader class of nonlinear drives, in other words a more general source for the $k_\alpha = 0$ component.  

Although we derive the source ($S$) of the residual from the initial condition, {\em i.e.} a delta function in time, the actual source in the gyrokinetic equation is the nonlinear term, which provides free energy to the $k_\alpha = 0$ component.  In the absence of a fully nonlinear theory describing the steady-state dynamics of the stable and unstable components of the turbulence, it is reasonable, in interpreting the result of the residual calculation, to consider what form a realistic turbulent source might take.  

One possibility is that the source has a significant non-zonal temperature perturbation, as expected from secondary instability theory \citep{Plunk-Banon_2017}, one mechanism by which zonal flows may be driven, which predicts that the temperature perturbation is both non-zonal $\delta T \neq \left< \delta T \right>$ and acquires its size and spatial dependence from the instabilities (ion-temperature-gradient) that drive it.  Other hints can be obtained directly from turbulence simulations.  Although the perpendicular temperature of the $k_\alpha = 0$ components is generally observed to be small \citep{Rogers-Dorland}, it is also observed to grow in relative amplitude at strong drive \citep{Plunk_2015}.  It is therefore unclear whether the temperature perturbations of our source should be expected to be large enough to drive a strongly non-zonal potential ($\delta T/(e_i\phi) \sim 1$), but it seems unlikely that they will always be so small that the non-zonal part of the residual can be assumed asymptotically small ($\delta T/(e_i\phi) \sim b_i$).  Similar questions also apply concerning other components of the source, such as parallel ion flow, and these will all have to be explored further in the future.

The work leaves ample opportunity for further studies, especially involving gyrokinetic simulations.  Fully nonlinear simulations of the turbulence may help identify cases where the non-zonal solutions described by Eqns.~\ref{eq:non-zonal-residual}-\ref{eq:full-residual} may arise.  On a more basic level, linear initial-value simulations should also be conducted to verify the quantitative validity of these expressions, especially for recently found designs that satisfy the quasi-isodynamic condition to high precision \citep{goodman2022}.  It should be noted that the predictions of this work apply also for the much simpler context of tokamak geometry. In particular, Eqn.~\ref{eq:res-zf} gives a prediction for the residual when the initial condition is non-zonal, which may already be tested for simple model tokamak geometries with spatially varying flux compression ($|\nabla \psi|$).  The inequality derived to bound the residual, Eqn.~\ref{eq:D-inequality-1}, might also be further investigated in some limits, and it, along with related estimates, could prove useful in stellarator optimization for reduced turbulence.

{\bf Acknowledgements.}  We thank P. Xanthopoulos for motivating this work, and thank Eduardo Rodr{\'i}guez, Iv{\'a}n Calvo and Felix Parra for helpful discussions.

{\bf Funding.}  This work has been carried out within the framework of the EUROfusion Consortium, funded by the European Union via the Euratom Research and Training Programme (Grant Agreement No 101052200 — EUROfusion). Views and opinions expressed are however those of the author(s) only and do not necessarily reflect those of the European Union or the European Commission. Neither the European Union nor the European Commission can be held responsible for them.  This work was partly supported by a grant from the Simons Foundation (560651, PH).

{\bf Declaration of Interests}. The authors report no conflict of interest.

{\bf Author ORCID.} G. G. Plunk, https://orcid.org/0000-0002-4012-4038. P. Helander, https://orcid.org/0000-0002-0460-590X.

\appendix

\section{Zonal flow oscillations in non-omnigenous stellarators}\label{appx:zf-oscillations}

The works of \citet{Mishchenko-Helander-Koenies, Helander_2011}, etc, identified zonal flow behavior special to (non-omnigenous) stellarators, involving the secular radial drifts; see also \citet{Monreal_2017} for generalizations.  To give a sense of how such oscillatory solutions arises in the present work, we consider equation \ref{eq:full-integral-eqn-2}, focusing on the non-resonant limit, with the specific ordering $k \overline{v}_{ra}/p \sim b_a^{1/2} \sim k_r\delta_{ra}$:

\begin{multline}
    \sum_a n_a\frac{e_a^2}{T_a}\left(\hat{\phi}  - \frac{1}{n_a} \int d^3v F_{a0} \overline{\hat{\phi}}\right) + \sum_a \frac{e_a^2}{T_a}\int d^3 v \frac{k_r^2 \overline{v}_{ra}^2}{p^2}F_{a0}\overline{\hat{\phi}} \\
+ \sum_a \frac{e_a^2}{T_a}\int d^3 v F_{a0}\left[\overline{\hat{\phi}} b_a x_\perp^2/2 + \overline{\hat{\phi} b_a x_\perp^2/2} + k_r^2 \overline{\hat{\phi}\delta_{ra}}\delta_{ra} - k_r^2 \overline{\hat{\phi}\delta_{ra}^2}- k_r^2 \overline{\hat{\phi}}\delta_{ra}^2\right] = \frac{S}{p}.
\end{multline}
By the same arguments made in the main text, we can consider $S$ such that the potential $\hat{\phi}$ is zonal to dominant order, and we can pull out a factor of the zonal potential $\hat{\Phi}$.  We need only consider the right hand side of the resulting equation whose roots (analytically continued) in the complex $p$-plane yield the damping rate and frequency of the modes of interest, in particular the imaginary part (giving the real frequency of the mode) is obtained from the zeros of

\begin{equation}
    \sum_a \frac{e_a^2}{T_a} \left(\int d^3v F_{a0} \frac{k_r^2\overline{v}_{ra}^2}{p^2} + \left< \int d^3v F_{a0}\left( b_a^2 x_\perp^2 + k_r^2(\delta_{ra}^2-\overline{\delta_{ra}}^2) \right) \right> \right).
\end{equation}

\section{Residual for general Larmor radius and orbit width}\label{appx:general-residual}
The limit $k\overline{v}_{ra} \ll p$ can be easily applied to equation \ref{eq:full-integral-eqn} to give an integral equation for the (`intermediate') residual potential without any assumptions about the size of $k_r\delta_{ra}$ or $k_\perp \rho_a$:

\begin{multline}
\sum_a \frac{e_a^2}{T_a}\left(n_a\phi_\mathrm{res} - \int d^3v J_{0a}F_{a0} \overline{\phi_\mathrm{res} J_{0a}e^{i k_r \delta_{ra}}} e^{-i k_r \delta_{ra}}\right)\\
    = \sum_a e_a \int d^3v J_{0a}\overline{\delta F_a(0) e^{i k_r \delta_{ra}}}e^{-i k_r \delta_{ra}},\label{eq:general-residual}
\end{multline}
An equation similar to this was given by \citet{Rosenbluth-Hinton-PhysRevLett.80.724} (see equation 8 there), where it was argued that a solution must exist due to the associated variational principle.  The expression can also be compared with the results of \cite{Monreal-2016}, specializing to the cases where the approximation $\phi_\mathrm{res} \approx \left< \phi_\mathrm{res} \right>$ is accurate.

\section{Ordering of the initial condition and the source term}\label{appx:deltaF0}

At zeroth order, and at order $\delta_r k_r \sim b_i^{1/2}$, quasi-neutrality for the initial condition is trivial,

\begin{align}
    \sum_a e_a \int d^3v \delta F_{a}^{(0)}(0) = 0,\\
    \sum_a e_a \int d^3v \delta F_{a}^{(1/2)}(0) = 0,
\end{align}
implying $\left<S_0\right> = \sum_a e_a \left<\int d^3v \overline{\delta F_{a}^{(0)}(0)}\right> = 0$, and $\left<S_{1/2}\right> = \sum_a e_a \left<\int d^3v \overline{\delta F_{a}^{(1/2)}(0)}\right> = 0$.  We note that this does not require either $\delta F_{a}^{(0)}(0)$ or $\delta F_{a}^{(1/2)}(0)$ to vanish, but strongly constrains their form.  At first order, we obtain

\begin{equation}
    \sum_a e_a \int d^3v \delta F_{a}^{(1)}(0) = n_i\frac{e_i^2}{T_i}b_i \phi(0) + e_i \int d^3v \delta F_{a}^{(0)}(0)b_ix_\perp^2/2
\end{equation}
Finally, averaging this first order constraint over a flux surface yields the terms that are needed to compute the source to second order, contributing to $\widetilde{\Phi}_S$:

\begin{equation}
    \sum_a e_a \left<\int d^3v \overline{\delta F_{a}^{(1)}(0)}\right> =  n_i\frac{e_i^2}{T_i}\left<b_i \phi(0)\right> + e_i \left<\int d^3v \delta F_{a}^{(0)}(0)b_ix_\perp^2/2\right>.\label{eq:initial-charge}
\end{equation}

Comparing with a case studied by \citet{Rosenbluth-Hinton-PhysRevLett.80.724}, we mention the possibility of retaining a nonzero $\delta F_{a}^{(1/2)}(0)$ which is odd in $v_\|$, and therefore consistent with quasi-neutrality being zero to this order, and also with the condition $S_0 = 0$, needed to neglect $\delta\phi_\mathrm{res}$ in equation \ref{eq:full-residual}.  This does however assume $\delta F_{a}(0)$ to be much larger than the typical ordering, which is linear in $b_i$.

\section{Useful identities}\label{appx:identities}
 First, a note on the notation: the infinite domain of the arc length variable $l$ can be divided into a set of intervals that we call ``wells''.  Each well consists of all the bounce points for the set of trapped particles with $1/B_\mathrm{max,j} < \lambda < 1/B_\mathrm{min,j}$.  Thus the integration over the domain is written as a sum of averages over such wells.  In the simple case where there is a single maximum and minimum of the magnetic field strength on the flux surface, this classification is straightforward, as each maximum marks the division between the wells.  For more complicated cases, the way of making the division is not uniquely determined, but it is straightforward to set the boundaries according to all the local maxima that occur along the field lines.  After this tedious task is done, the summation over all wells includes all points along the field line.  In the sum over well indices, it is convenient to reserve the $j=0$ `well' as the domain of the passing particles, {\it i.e.} the entire interval $(-L,L)$ over which the limit is taken $L \rightarrow \infty$.  Thus for $\lambda < 1/B_\mathrm{max}$, with $B_\mathrm{max}$ the global maximum of $B(l)$, the bounds of the transit average are $(l_1,l_2) = (-L, L)$, and the average is understood in the limiting sense as $L \rightarrow \infty$.
 
 Following the above discussion, it is possible to exchange the order of integration over the field line with integration over the phase space variable $\lambda$ as follows:
\begin{equation}
    \int_{-L}^{L}\frac{dl}{B} \int \frac{B d\lambda}{\sqrt{1-\lambda B(l)}} = \sum_j \int d\lambda \int_{l_1(\lambda, j)}^{l_2(\lambda, j)} \frac{dl}{\sqrt{1-\lambda B}}
\end{equation}
From this identity it is straightforward to derive the following
\begin{equation}
    \left< \int d^3 v \overline{f} \right> = \lim_{L\rightarrow \infty}\frac{1}{V} \int_{-L}^{L}\frac{dl}{B} \int d^3 v \overline{f} = \left< \int d^3 v f \right>,\label{eq:flux-bounce-avg}
\end{equation}
where $V = \int_{-L}^{L}dl/B$, and we have used

\begin{equation}
    d^3v = \sum_{\sigma} \frac{\pi B v^2 d\lambda dv}{\sqrt{1-\lambda B}}.
\end{equation}

\section{Positivity of ${\cal L}$}\label{appx:positivity}
The operator ${\cal L}$ can be defined, using $F_{a0} = n_a \exp(-v^2/v_{Ta}^2)/(v_{Ta}^3\pi^{3/2})$ and $v_{Ta} = \sqrt{2T_a/m_a}$, as

\begin{equation}
    {\cal L} \phi \equiv \phi  - \frac{B}{2}\int_0^{1/B} \frac{d\lambda}{\sqrt{1-\lambda B}}  \overline{\phi}.
\end{equation}
Multiplying this equation by $\phi^*$, integrating over the flux surface, and using the identity \ref{eq:flux-bounce-avg} we obtain

\begin{align}
    \int \frac{dl}{B} \phi^* {\cal L}\phi &= \int \frac{dl}{B} \phi^*\left( \phi  - \frac{B}{2}\int_0^{1/B} \frac{ \overline{\phi} \; d\lambda}{\sqrt{1-\lambda B}} \right)\\
    &= \frac{1}{2}\sum_j \int_0^{1/B}  \tau_j\left(\overline{|\phi|^2} - |\overline{\phi}|^2\right)d\lambda,
    \label{quadratic form}
\end{align}
where $\tau_j$ denotes the quantity \ref{bounce time} for the $j$'th trapping well. Because of the Schwarz inequality, $\overline{|\phi|^2} - |\overline{\phi}|^2 \geq 0$, Eqn.~\ref{quadratic form} is always greater or equal to zero, with equality only for the case that $\phi = \overline{\phi}$ for all $l$, {\em i.e.} $\phi = \left<\phi\right>$.  This is what was already shown by \citet{helander-microinstabilities-1}, but with the trivial modification of including passing part of phase space $\lambda < 1/B_\mathrm{max}$.
\bibliographystyle{unsrtnat}
\bibliography{residual-remix}

\end{document}